\documentclass[prl,twocolumn]{revtex4-1}

\usepackage{graphicx}
\usepackage{amsmath}
\usepackage{amssymb}
\usepackage{dcolumn}

\begin{document}

\newcommand{\dmskcomment}[1]{\marginpar{$\spadesuit$}}
\newcommand{\hami}{\mathcal{H}}

\newcommand{\nhat}{\hat{n}}
\newcommand{\nhatp}{\hat{n}_+}
\newcommand{\nhatm}{\hat{n}_-}
\newcommand{\nbarp}{\bar{n}_+}
\newcommand{\nbarm}{\bar{n}_-}
\newcommand{\nmaxpm}{\bar{n}_{\mbox{\scriptsize{max}},\pm}}
\newcommand{\nmax}{\bar{n}_{\mbox{\scriptsize{max}}}}
\newcommand{\chat}{\hat{c}}
\newcommand{\cdag}{\hat{c}^\dagger}
\newcommand{\cphat}{\hat{c}_+}
\newcommand{\cpdag}{\hat{c}^\dagger_+}
\newcommand{\cmhat}{\hat{c}_-}
\newcommand{\cmdag}{\hat{c}^\dagger_-}
\newcommand{\omc}{\omega_c}
\newcommand{\dca}{\Delta_{ca}}
\newcommand{\dn}{\Delta_N}
\newcommand{\delp}{\Delta_p}
\newcommand{\delppm}{\Delta_{p,\pm}}
\newcommand{\nbar}{\bar{n}}
\newcommand{\nbarmax}{\bar{n}_{\mbox{\scriptsize{max}}}}
\newcommand{\Omc}{\Omega_c}
\newcommand{\Omeff}{\mathbf{\Omega}_{\mbox{\scriptsize{eff}}}}
\newcommand{\Omeffrot}{\mathbf{\Omega}_{\mbox{\scriptsize{eff,r}}}}


\newcommand{\Oml}{\mathord{\Omega_L}}
\newcommand{\Omlp}{\mathord{\Oml^\prime}}
\newcommand{\Omlpp}{\mathord{\Oml^{\prime \prime}}}
\newcommand{\Shat}{\hat{S}}
\newcommand{\shat}{\hat{s}}
\newcommand{\Shatbf}{\mathbf{\hat{S}}}
\newcommand{\shatbf}{\mathbf{\hat{s}}}
\newcommand{\Sbf}{\mathbf{S}}
\newcommand{\Sbfrot}{\mathbf{S}_{\mbox{\scriptsize{r}}}}
\newcommand{\beff}{\mathbf{B}_{\mbox{\scriptsize{eff}}}}
\newcommand{\bax}{\mathbf{b}}
\newcommand{\ssql}{\Delta S_{\mbox{\scriptsize{SQL}}}}

\newcommand{\zhat}{\hat{z}}
\newcommand{\phat}{\hat{p}}
\newcommand{\omz}{\omega_z}
\newcommand{\ahat}{\hat{a}}
\newcommand{\adag}{\hat{a}^\dagger}
\newcommand{\zho}{z_{\mbox{\scriptsize HO}}}
\newcommand{\pho}{p_{\mbox{\scriptsize HO}}}

\newcommand{\kax}{\mathbf{k}}
\newcommand{\iax}{\mathbf{i}}
\newcommand{\jax}{\mathbf{j}}
\newcommand{\kaxp}{\mathbf{k}^\prime}
\newcommand{\iaxp}{\mathbf{i}^\prime}
\newcommand{\jaxpr}{\mathbf{j}_{\mbox{\scriptsize{r}}}}
\newcommand{\kaxpr}{\mathbf{k}_{\mbox{\scriptsize{r}}}}
\newcommand{\iaxpr}{\mathbf{i}_{\mbox{\scriptsize{r}}}}
\newcommand{\jaxp}{\mathbf{j}^\prime}
\newcommand{\omrec}{\omega_{rec}}

\title{A spin optodynamics analogue of cavity optomechanics}
\author{N. Brahms$^1$ and D.M. Stamper-Kurn$^{1,2}$}
\email{dmsk@berkeley.edu} \affiliation{
    $^1$Department of Physics, University of California, Berkeley CA 94720, USA \\
    $^2$Materials Sciences Division, Lawrence Berkeley National Laboratory, Berkeley, CA 94720, USA}
\date{\today}%

\begin{abstract}
The dynamics of a large quantum spin coupled parametrically to an
optical resonator is treated in analogy with the motion of a
cantilever in cavity optomechanics.  New spin optodynamic phenomena
are predicted, such as cavity-spin bistability, optodynamic
spin-precession frequency shifts, coherent amplification and damping
of spin, and the spin optodynamic squeezing of light.
\end{abstract}

\maketitle

Cavity optomechanical systems are currently being explored with the
goal of measuring and controlling mechanical objects at the quantum
limit, using interactions with light \cite{kipp08sciencereview}.  In
such systems, the position of a mechanical oscillator is coupled
parametrically to the frequency of cavity photons.  A wealth of
phenomena result, including quantum-limited measurements
\cite{brag95qmbook}, mechanical response to photon shot noise
\cite{murc08backaction}, cavity cooling \cite{vule00}, and
ponderomotive optical squeezing \cite{note:pondero}.

Concurrently, spins and psuedospins coupled to electromagnetic
cavities are being researched in atomic \cite{haro06book}, ionic
\cite{hers09}, and nanofabricated systems \cite{barc09nv,dica09},
with applications including magnetometry \cite{gere03kalman}, atomic
clocks \cite{lero10squeeze}, and quantum information processing
\cite{turc95phase,haro06book,dica09}. In contrast to mechanical
objects, spin systems are more easily disconnected from their
environment and prepared in quantum states, including squeezed
states \cite{lero10squeeze}.

In this Rapid Communication, we seek to link these two fields by exploiting the
similarities between large-spin systems and harmonic oscillators
\cite{hols40} to construct a cavity spin optodynamics system in
analogy to cavity optomechanics.  Optomechanical phenomena map
directly to our proposed system, resulting in spin cooling and
amplification \cite{brag67,coolexpts}, nonlinear spin sensitivity
and spin-cavity bistability
\cite{dors83bistability,gupt07nonlinear}, and spin opto-dynamic
squeezing of light \cite{brag67,kimb01ligo}. Such a system may find
application as a quantum-limited spin amplifier or as a latching
spin detector. We detail these phenomena using currently accessible
parameters, and we propose realizations either using cold atoms and
visible light or using cryogenic solid state systems and microwaves.

An ideal cavity optomechanics system, consisting of a harmonic
oscillator coupled linearly to a single-mode cavity field, obeys the
Hamiltonian
\begin{equation} \hami = \hbar \omc \nhat + \hbar \omz
\adag \ahat  - f \zho \left( \adag + \ahat\right) \nhat +
\hami_{in/out}\label{eq:optohami}.
\end{equation}
Here $\ahat$ is the oscillator's phonon annihilation operator,
$\nhat$ is the photon number operator, $\omz$ is the natural
frequency of the oscillator in the dark, and $\omc$ is the bare
cavity resonance frequency.  $f$ is the radiation-pressure force
applied by a single photon, while $\zho = \sqrt{\hbar/2 m \omz}$ is
the harmonic oscillator length for oscillator mass $m$.
$\hami_\mathrm{in/out}$ describes the coupling of the cavity field
to external light modes.  Under this Hamiltonian, the cantilever
position $\zhat$ and momentum $\phat$ evolve as $d \zhat/d t =
\phat/m$ and $d \phat/d t = - m \omz^2 \zhat + f \nhat$.

\begin{figure}
\begin{centering}
\includegraphics[width=0.35\textwidth]{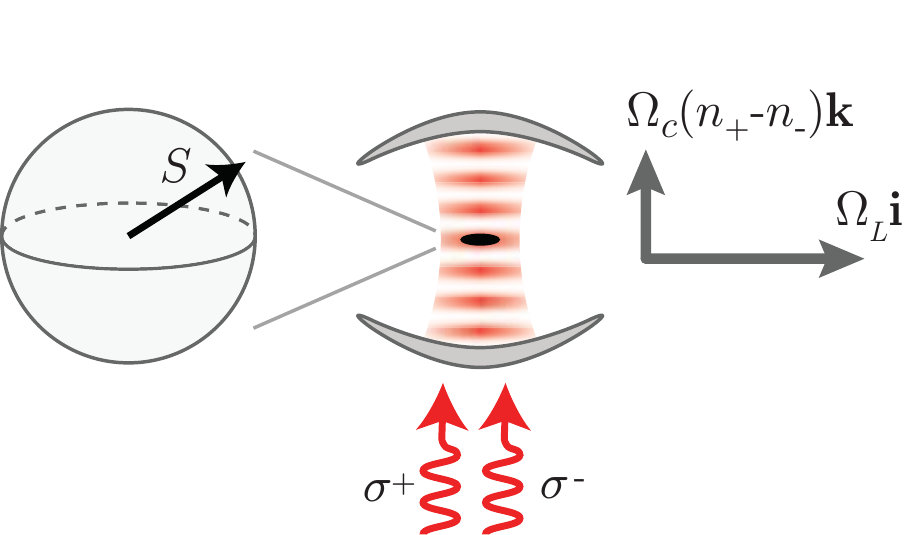}
\caption{(Color) An ensemble of atoms trapped within a driven
optical resonator experiences an externally imposed magnetic field
along $\iax$ and a light-induced effective magnetic field along the
cavity axis $\kax$.  The evolution of the collective spin $\Shatbf$
resembles that of a cantilever in cavity
optomechanics.}\label{fig:scheme1}
\end{centering}
\end{figure}

To construct a spin analogue of this system, we consider a
Fabry-Perot cavity with its axis along $\kax$ (Fig.\
\ref{fig:scheme1}).  For the collective spin, we first consider a
gas of $N$ hydrogenlike atoms in a single hyperfine manifold of
their electronic ground state, each with dimensionless spin $s$ and
gyromagnetic ratio $\gamma$.  The atoms are optically confined at an
antinode of the cavity field.  An external magnetic field
$\mathbf{B} = B \bax$ is applied to the atoms.  The detuning $\dca$
between the cavity resonance is chosen to be large compared to both
the natural linewidth and the hyperfine splitting of the atoms'
excited state.  In this limit, spontaneous emission may be ignored
and the single-atom cavity-field  interaction energy, $
\hami_\mathrm{Stark} = (\hbar g_0^2/\dca) \nhat \left ( 1 \pm
\upsilon \kax \cdot \shatbf \right )$, comprises the scalar and
vector ac Stark shifts \cite{happ67}, where $g_0$ quantifies the
atom-cavity coupling and  $\upsilon$ the vector shift.

Summing over all atoms $q$, we obtain the system Hamiltonian:
\begin{multline}
  \hami = \hbar \omc \left(\nhatp +\nhatm\right) +  \hami_{in/out} +  \sum_q\! \bigg(\!  -\hbar \gamma \mathbf{B} \cdot \shatbf_q \\
+ \frac{\hbar g_0^2}{\dca} \left[\left(\nhatp + \nhatm \right) +
\upsilon \left(\nhatp - \nhatm \right) \kax \cdot \shatbf_q \right]
\bigg),
\end{multline}
with number operators $\nhat_\pm$ for the $\sigma^\pm$ polarized
optical modes.

The above Hamiltonian can be rewritten as the interaction of the
collective spin operator $\Shatbf \equiv \sum_q \shatbf_q$ with an
effective total magnetic field $\beff \equiv  \Omeff / \gamma$,
giving \cite{cct72}
\begin{equation}
\Omeff = \Oml \bax + \Omc \left(\nhatp - \nhatm \right) \kax.
\end{equation}
Here $\Oml = \gamma B$ and $\Omc = - \upsilon g_0^2 / \dca$.
Altogether, the cavity spin optodynamical Hamiltonian is
\begin{equation}
\hami = \hbar \left (\omc + \frac{N g_0^2}{\dca} \right ) \left(\nhatp + \nhatm \right)
+\hami_{in/out} - \hbar \Omeff \cdot \mathbf{\Shat}.
\label{eq:sodhami}
\end{equation}

Now consider the external magnetic field to be static and oriented along $\iax$, orthogonal to the cavity axis. In the limit $\langle\Shat\rangle \simeq S \iax$, the spin dynamics become
\begin{align}
\frac{d \Shat_j}{d t} &= \Oml \Shat_k -\Omc S \left( \nhatp \mathord{-}
\nhatm \right), & \frac{d \Shat_k}{d t} &= -\Oml \Shat_j.
\label{eq:limitsyzdot}
\end{align}
The analogy between cavity optomechanics and spin optodynamics is
established by assigning $\zhat$~$\rightarrow$~$- \zho
\Shat_k/\ssql$ and $\phat$~$\rightarrow$~$\pho \Shat_j / \ssql$,
where $\zho$ and $\pho = \hbar / (2 \zho)$ are defined with $\omz
\!\rightarrow\! \Oml$ \cite{hols40} and $\ssql=\sqrt{S/2}$ is the
standard quantum limit for transverse spin fluctuations.
Eqs.~(\ref{eq:limitsyzdot}) now match the optomechanical equations
of motion with the optomechanical coupling defined through $ f \zho
\nhat \!\rightarrow\! - \hbar \Omc \ssql (\nhatp \mathord{-}
\nhatm)$. The main result of this work, that various cavity
optomechanical phenomena are manifest also in cavity spin
optodynamical  systems, is immediately established.

Let us now elaborate on these phenomena. To obtain general results,
we will proceed without assuming $\Shat \simeq S\iax$, except in
certain cases, noted in the text, where some physical insight is
gained.  We begin with effects for which both the light field and
the ensemble spin may be treated classically, i.e.\ by letting $\Sbf
= \langle \Shatbf\rangle$ and $\bar{n}_\pm = \langle \hat{n}_\pm
\rangle$.

\begin{figure}
\begin{centering}
\includegraphics[width=0.45\textwidth]{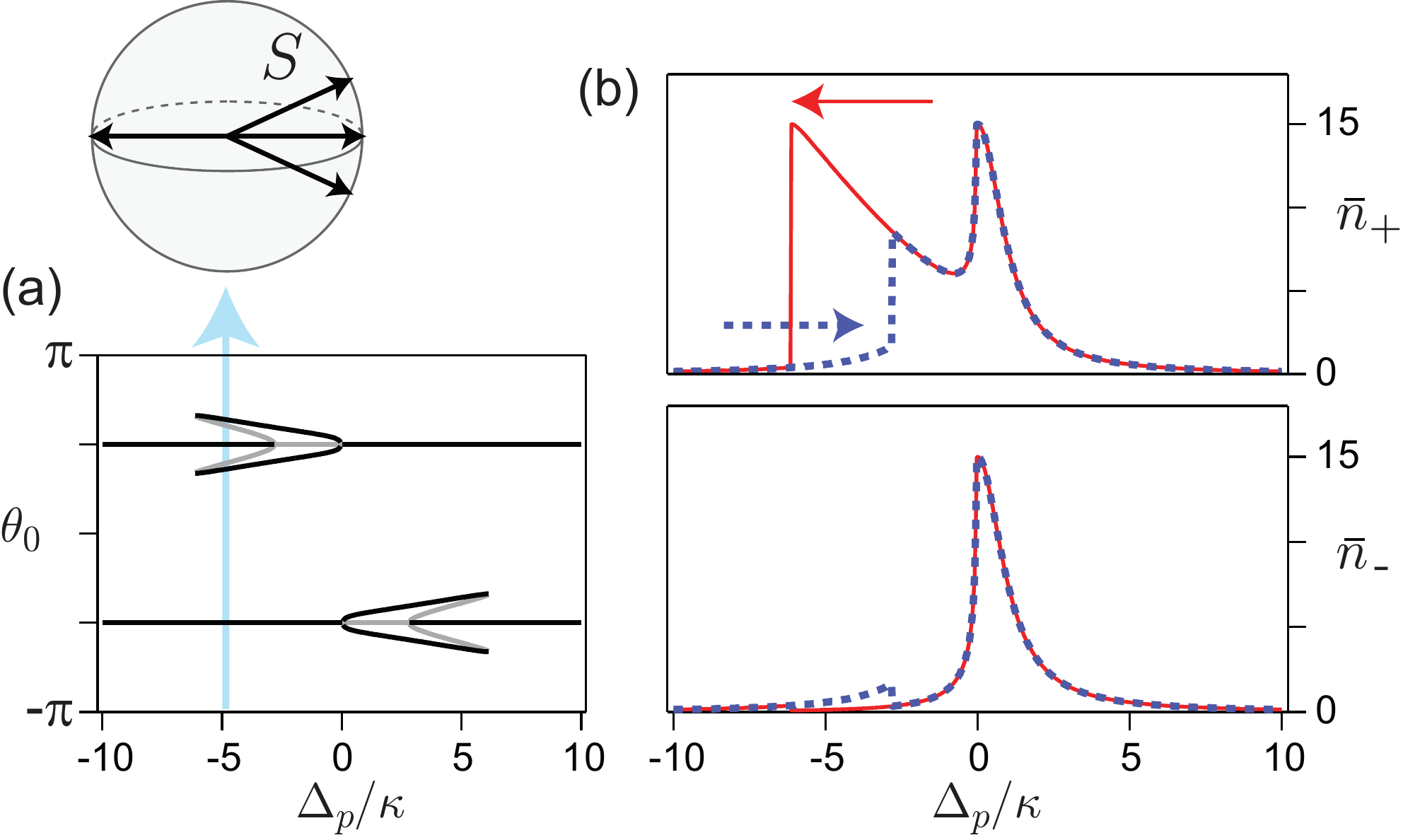}
\caption{(Color) Cavity-spin bistability in a cavity driven with
linearly polarized light.  We consider  $N=5000$ spin-2 $^{87}$Rb
atoms, $\Omc/\kappa = 1.25 \times 10^{-3}$, $\Oml/\kappa = 3.3
\times 10^{-2}$, and $\nbar_{max,\pm} = 15$ (similar to Ref.\
\cite{purd10tunable}). (a) As $\Delta_p$ is varied, several stable
(black) and unstable (gray) static spin configurations are found.
Configurations for $\Delta_p/\kappa = -4.8$ are depicted.  (b) The
cavity exhibits hysteresis as the probe is swept with positive
(dashed blue) or negative (red) frequency chirps, with the spin
initially along $\iax$. Rapid transitions as $\Delta_p/\kappa$ is
swept upward from -2.8 or downward from 0 involve symmetry breaking
as the cavity becomes birefringent; we display $\nbar_+$ and
$\nbar_-$ assuming the stable branch closer to $\theta_0=0$ is
selected. Here, $\dca/2\pi = 20$ GHz from the D2 transition, $g_0/2
\pi = 15$ MHz, $\kappa/2\pi = 1.5$ MHz.} \label{fig:bistability}
\end{centering}
\end{figure}

\textbf{Cavity-spin bistability:}
We start with the static behavior of the system by finding the fixed points of the system.  The collective spin vector is static when $\Sbf$ is parallel to $\Omeff$.  Writing $\Sbf = S ( \iax \sin\theta_0 + \kax \cos \theta_0)$, this condition requires $\nbarp-\nbarm = (\Oml/\Omc) \cot \theta_0$. The intracavity photon numbers are determined also by the standard expression for a driven cavity of half line-width $\kappa$, i.e.\ $ \bar{n}_\pm = \nmaxpm [1 + \left(\delppm \pm \Omc S \cos\theta_0 \right)^2 / \kappa^2]^{-1}$ with $\omega_\pm = (\omc + N g_0^2/\dca)+\delppm$ being the frequency of laser light of polarization $\sigma^\pm$ driving the cavity and $\nmaxpm$ characterizing its power.  These two expressions for $\nbarp-\nbarm$ may admit several solutions (Fig.\ \ref{fig:bistability}).

As typical in instances of cavity bistability \cite{gibb85book}, several of the static solutions for the intracavity intensities may be unstable.  To identify such instabilities, we consider the torque on the collective spin when it is displaced slightly toward $+\kax$ from its static orientation.  Stable dynamics result when such displacement yields a torque $\mathbf{N} \cdot \jax$ with the sign $\alpha = \textrm{sgn}(\sin\theta_0)$. Geometrically, this stability requires that the spin vector be displaced further in the $+ \kax$ direction than the vector $\alpha \Omeff$. Quantifying the linear response of the intracavity effective magnetic field to variations of the collective spin via $\lambda = \Omc d (\nbarp - \nbarm)/d S_k$, the static spin orientations are found to be unstable when $\alpha \lambda > \Oml |\csc^3 \theta_0|/ S$.

\begin{figure}
\begin{center}
\includegraphics[width=0.45\textwidth]{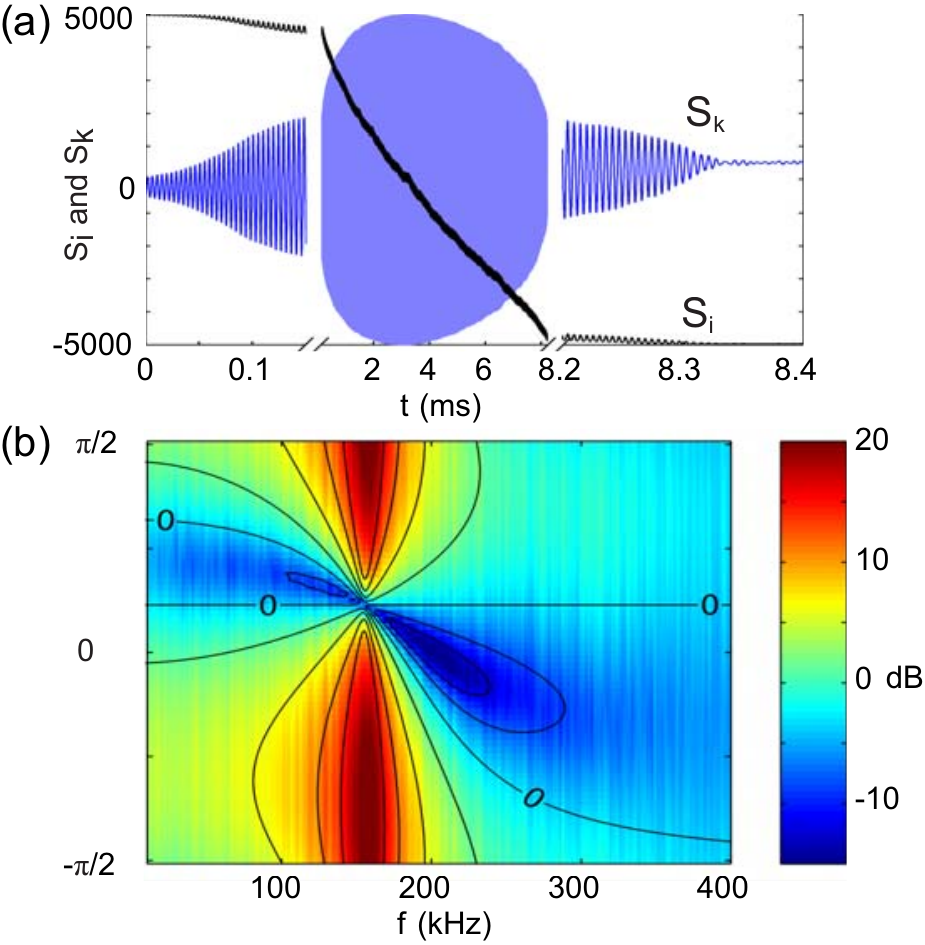}
\end{center}
\caption{(Color) Simulations of spin dynamics for $S=5000$,
$\Oml/2\pi = 200$~kHz, $\Omc/2\pi = -2.3$~kHz, $\kappa/2\pi =
1.8$~MHz, $\nhatp = 10$, and  $\Delta_{p,+}= 0.37 \kappa$. (a) Time
evolution of $S_i$ (black) and $S_k$ (blue), following spin
preparation near $\iax$, shows amplification, reorientation, and
damping toward the high-energy stable orientation near $-\iax$. Note
the different scales on the horizontal axis.  (b) Logarithmic
optical spectral noise power relative to that of coherent light,
plotted vs.\ quadrature angle $\phi$ (amplitude quadrature at
$\phi=0$), shows inhomogeneous optical squeezing. Simulation results
shown in color, and linearized theory (Eq.\ \ref{eq:lintheory}) as
contour lines every 5 dB.} \label{fig:simulation}
\end{figure}

\textbf{Opto-dynamical Larmor frequency shift:} The dynamics of the
spin precessing about one of the stable configurations can be
parameterized by the precession frequency, which is shifted from
$\Oml$ by two effects. First, there is an upward frequency shift
from the static modification of the effective magnetic field,
leading to precession at the frequency $\Oml^\prime = \Oml
|\csc\theta_0|$ when $\lambda = 0$. A second shift occurs when the
spin dynamics are slow compared to the response time of the cavity
field ($\Oml^\prime \ll \kappa$). Here, the precessing spin
modulates the cavity field, which, in turn, acts back upon the spin
to modify its precession frequency, When the precession amplitude is
small, a solution of the spin equations of motion derived from the
Hamiltonian in Eq.~(\ref{eq:sodhami}) yields an overall precession
frequency $\Omlpp$, where
\begin{equation}
\left. \Omlpp \right.^2 = {\Omlp}^2 - \lambda \Oml S \sin \theta_0.
\label{eq:optoshift}
\end{equation}
The quantity $k_S \equiv - \lambda \Oml S \sin\theta_0$ serves as
the analogue of the optical spring constant \cite{buon02spring}, and
leads to shifts of the Larmor precession frequency with a sign and
magnitude that depend on the spin orientation, $\lambda$ and the
frequency, intensity, and polarization of the cavity probe fields.
When the precession amplitude is large, the dynamics become
essentially nonlinear.  In this case the dynamics can be described
by numerical simulation (Fig.~\ref{fig:simulation}).

\textbf{Coherent amplification and damping of spin:} Now we consider
the effects of the finite cavity response time $\kappa^{-1}$ on the
spin dynamics. To develop an intuitive picture, we consider the
unresolved sideband regime $\Oml < \kappa$, in a frame (indicated by
the index ``r'') corotating with the collective spin, with $\iaxpr$
aligned to the fixed point. We assume the spin to be precessing at a
near constant rate, and the cavity field response to this precession
to be simply delayed by $\kappa^{-1}$. Employing the rotating-wave
approximation, the delay causes the effective field $\Omeffrot$ to
point out of the $\iaxpr$-$\kaxpr$ plane, with $\Omeffrot\cdot\jaxpr
= -(\alpha \lambda S_{k,r} \sin^2\theta_0 \sin\phi)/2$, where $\phi
= \Omlpp /\kappa$. The collective spin now experiences a torque in
the $\kaxpr$ direction, giving
\begin{equation}
\frac{d S_{k,r}}{d t} = \frac{-\alpha \lambda \sin^2\theta_0
\sin{\phi} S_{i,r}}{2} S_{k,r}
\label{eq:optodamp}
\end{equation}
For positive (negative) values of $\alpha \lambda$ , the Larmor
precession frequency is shifted down (up) and the spin is damped
toward (amplified away from) its stable point. Similar relations
apply to cavity optomechanics \cite{corb07gram}. The deflection of
the spin toward or away from the stable points persists for large
precession amplitudes (Fig.~\ref{fig:simulation}).

This cavity-induced spin amplification or damping differs from
conventional optical pumping in two important respects.  First,
while the spin polarization generated by optical pumping relies on
the polarization of the pump light, the target state for
cavity-induced spin damping is selected energetically.  Similar to
cavity optomechanical cooling \cite{vule00}, cavity enhancement of
Raman scattered light drives spins to the high- or low-energy spin
state according to the detuning of probe light from the cavity
resonance, independent of the polarization. Second, this
amplification or damping of the intracavity spin is coherent, preserving the phase of Larmor precession, at least within the limits of a quantum amplifier.

\textbf{Spin optodynamical squeezing of light:} We now consider
quantum optical effects of cavity spin optodynamics. One such effect
is the disturbance of the collective spin due to quantum optical
fluctuations of the cavity fields.  In cavity optomechanics,
intracavity photon number fluctuations disturb the motion of a
cantilever, providing the necessary backaction of a quantum
measurement of position \cite{cave80theydo}. The analogous
disturbance of optically probed atomic spins (or pseudo-spins) has
been studied both in free-space \cite{scho02spin} and intracavity
\cite{schl10states,lero10squeeze} configurations. In an
optomechanics-like configuration, e.g.\ with $\mathbf{B}\propto
\iax$, backaction heating of the atomic spin enforces quantum limits
to measurement of the precessing ensemble and also set limits on
optodynamical cooling.  In contrast with optomechanical systems,
optically probed spin ensembles readily present the opportunity to
perform quantum-non-demolition (QND) measurements; with
$\mathbf{B}\propto\kax$, the detected spin component $S_k$ is a QND
variable representing the energy of the spin system.

The noise-perturbed spin acts back upon the cavity optical field,
mediating a self-interaction of the light field that can result in
optical squeezing. To exhibit this effect, we consider a cavity
illumined with $\sigma^+$ circular polarized probe light with
detuning $\Delta_{p}$.  The dynamics of the cavity field are given
by
\begin{equation}
\frac{d \hat{c}_+}{d t} = (i \Delta_{p} - \kappa + i \Omc \Shat_k)
\hat{c}_+ + \kappa \left(\eta + \hat{\xi}_+\right).
\end{equation}
Here, $\eta$ gives the coherent-state amplitude of the drive field
and the noise operator $\hat{\xi}_+$ represents its fluctuations.  When evaluating the dynamics numerically, we consider a
semiclassical Langevin equation, converting $\hat{\xi}_+$ into a
Gaussian stochastic variable with statistics related to those of the
noise operator, and replacing the operators $\hat{c}_+$ and
$\Shatbf$ with $c$-numbers. This substitution is appropriate for
moderately large values of $\nbar$ and $S$.

Fig.~\ref{fig:simulation} portrays the simulated evolution of a
spin prepared initially in a low-energy spin orientation (close to
$\iax$), driven by a blue-detuned cavity probe. Coherent spin
amplification directs the spin toward the stable high-energy
configuration (near $-\iax$), yielding a dynamical steady state
characterized by a negative temperature.

To obtain analytical expressions for the evolution dynamics, we
follow the example of cavity optomechanics \cite{note:pondero} by
linearizing the Langevin equations for spin and optical fluctuations
about their steady-state value.  The spin projection $S_k$ responds
to amplitude-quadrature fluctuations of the cavity field
$\xi_A(\omega)$ with susceptibility
\begin{equation}
\chi(\omega) \equiv \frac{S_k(\omega)}{\xi_A(\omega)} = \frac{-
\Omlp \Omega_c \sqrt{\nbar_+}}{\Omlp^2 + k_S R(\omega) - \omega^2 + i
\omega \Gamma_o (\omega)},
\end{equation}
where $\Gamma_o(\omega) = 2 \kappa \frac{\Omlp^2 -
\omega^2}{\kappa^2 + \Delta_p^2 - \omega^2}$ is the cavity
optodynamic spin damping, and $R(\omega) =
\frac{\kappa^2+\Delta_p^2}{\kappa^2 + \Delta_p^2 - \omega^2}$.  The
susceptibility is largest for $\omega \simeq \Omlpp$. The driven
spin feeds the fluctuations back onto the cavity field, yielding the
intracavity field fluctuation spectrum
\begin{equation}
c_+(\omega) = \frac{ \Omlp^2 + i \frac{\kappa+\omega}{\Delta_p}k_S R(\omega) - \omega^2 + i \omega \Gamma_o(\omega) }
{\Omlp^2 + k_S R(\omega) - \omega^2 + i \omega \Gamma_o(\omega)} \, \xi_A + i
\xi_P,
\label{eq:lintheory}
\end{equation}
where $\xi_P(\omega)$ is the input spectrum of phase fluctuations.
This fluctuation spectrum exhibits inhomogeneous optical squeezing
(Fig.\ \ref{fig:simulation}b).

\textbf{Applications:} The analogy of cavity optodynamics widens the
range of phenomena accessed through the manipulation and detection
of quantum spins within optical cavities, enabling several
applications.  For example, bistability in cavity-coupled
single-spin systems serves to increase the readout fidelity of
cavity-coupled qubits \cite{sidd06disp}. Similarly here, cavity-spin
bistability could be used as a Schmitt trigger for the collective
spin:  if the probe power is turned on diabatically in the bistable
regime, the cavity transmission will latch into either a bright or
dark state, depending on whether the initial spin state is below or
above a parametrically chosen threshold value.  The cavity-spin
system may also be used as a phase-preserving amplifier for spin
dynamics occurring near the shifted precession frequency $\Omlpp$,
with amplification noise given in Eq.~(\ref{eq:lintheory}). Both
applications may aid measurements of ac magnetic fields, amplifying
weak signals above technical sensitivity limits.

Conversely, cavity spin optodynamics may be applied as a powerful
simulator of cavity optomechanics, with the spin system allowing for
new means of control. For example, precession frequencies may be
tuned rapidly by varying the applied magnetic field, simulating
optomechanics with a dynamically variable mechanical spring
constant.  Alternately, spatial control of inhomogeneous magnetic
fields may be used to divide a spin ensemble into several
independent subensembles, simulating optomechanics with several
mechanical modes.

In addition to the dilute gas implementation discussed so far, a
similar system could be constructed using solid-state spin ensembles
and microwave resonators.  For example, using ensembles of
nitrogen-vacancy defects in diamond coupled to the circular
polarized evanescent radiation of a crossed microwave resonator
\cite{hend08}, the Hamiltonian of Eq.~(\ref{eq:sodhami}) is obtained
by the ground $m_s = \pm 1$ electronic states as the pseudospin and
replacing the ac Stark shift with an ac Zeeman shift from microwave
radiation near the 2.8-GHz crystal-field-split Zeeman transition.

We thank H.\ Mabuchi and K.B.\ Whaley for inspiring discussions.
This work was supported by the NSF and the AFOSR. D.M.S.-K.
acknowledges support from the Miller Institute for Basic Research in
Science.

\end{document}